\documentstyle[12pt]{article}
\begin{document}
\begin{titlepage}
\begin{flushright}
LNF--00/013(P)\\
UAB--FT/485\\
hep-ph/0003273\\
March 2000
\end{flushright}
\vspace*{1.6cm}

\begin{center}
{\Large\bf The ratio $\Phi\rightarrow K^+K^-/K^0\bar K^0$}\\
\vspace*{0.8cm}

A.~Bramon$^1$, R.~Escribano$^2$, J.L.~Lucio M.$^{2,3}$, and
G.~Pancheri$^2$\\
\vspace*{0.2cm}

{\footnotesize\it $^1$Departament de F\'{\i}sica,
Universitat Aut\`onoma de Barcelona, 
E-08193 Bellaterra (Barcelona), Spain\\
$^2$INFN-Laboratori Nazionali di Frascati,
P.O.~Box 13, I-00044 Frascati, Italy\\
$^3$Instituto de F\'{\i}sica, Universidad de Guanajuato,
Lomas del Bosque \# 103, Lomas del Campestre, 37150 Le\'on, 
Guanajuato, Mexico}
\end{center}
\vspace*{1.0cm}

\begin{abstract}
The ratio $\Phi\rightarrow K^+K^-/K^0\bar K^0$ is discussed and its 
present experimental value is compared with theoretical expectations. 
A difference larger than two standard deviations is observed. 
We critically examine a number of mechanisms that could 
account for this discrepancy, which remains unexplained. 
Measurements at DA$\Phi$NE at the level of the per mille accuracy can 
clarify whether there exist any anomaly.
\end{abstract}
\end{titlepage}

\section{Introduction}

The $\phi$-meson was  discovered many years ago as a $K \bar K$ 
resonance \cite{discovery}.  
Its decay is dominated by the two $K \bar K$ decay modes which proceed 
through Zweig-rule allowed strong interactions. The ratio
$R \equiv \phi \to K^+ K^- / K^0 \bar{K^0}$ has been measured in a variety
of independent experiments using  different $\phi$-production mechanisms. 
Among these, the cleanest one is electron-positron annihilation around the 
$\phi$ resonance peak, 
{\it i.e.}~the reactions $e^+e^-\to \phi \to K^+ K^-/K^0 \bar{K^0}$, 
which have been accurately measured at Novosibirsk quite recently \cite{AKH95} 
and are the object of intense investigation at the Frascati 
$\Phi$-factory \cite{daphne}. 
With as much as $8\times 10^6\ \phi$'s on tape, the KLOE experiment at 
DA$\Phi$NE can be expected to measure the above ratio $R$ with a statistical
accuracy of the order of the per mille. In view of this, we wish to discuss 
the theoretical expectations and compare them with the most recent 
determinations for this ratio. 

In the following we shall first review the present experimental situation,
then compare it with the na\"{\i}ve expectations from isospin symmetry and
phase space considerations thus observing that a disagreement seems to
exist.
Contributions arising from electromagnetic radiative
corrections and $m_u-m_d$ isospin breaking effects are analyzed and shown 
to bring the observed discrepancy to be more than three standard
deviations. 
Various additional theoretical improvements on our analysis, such as the 
use of vector-meson dominated electromagnetic form-factors, 
the modification of the strong vertices and the inclusion 
of rescattering effects through the scalar resonances $f_0(980)$ and 
$a_0(980)$ using the charged kaon loop model, are also examined and shown
not to change in any substantial way our results which imply a clear
discrepancy between theory and the available data.

%\section{Experimental data}

The first combined measurement of the four major $\phi$ decay modes in a 
single $e^+ e^-$ dedicated experiment has been performed quite recently 
with the general purpose detector CMD-2 at the upgraded $e^+ e^-$ collider
VEPP-2M at Novosibirsk \cite{AKH95}. 
Having a single experiment normalized to almost 100\% of decay modes
implies a reduction of systematic errors, and the following branching ratios 
(BR) and errors from VEPP-2M \cite{AKH95} are quoted: 
\begin{equation}
\begin{array}{l}
{\rm BR}(\phi\rightarrow K^+K^-)=(49.2\pm 1.2)\%\ ,\\[2ex]
{\rm BR}(\phi\rightarrow K^0\bar K^0)=(33.5\pm 1.0)\%\ ,
\end{array}
\end{equation}
leading to 
\begin{equation} 
R_{\rm exp}\equiv
\frac{{\rm BR}(\phi\rightarrow K^+K^-)}
     {{\rm BR}(\phi\rightarrow K^0\bar K^0)}=1.47\pm 0.06\ .
\end{equation}
All these results were in agreement with the average values quoted in the 
then available PDG 1994 compilation \cite{pdg94}:
\begin{equation}
\left.
\begin{array}{l}
{\rm BR}(\phi\rightarrow K^+K^-)=(49.1\pm 0.9)\%\\[2ex]
{\rm BR}(\phi\rightarrow K^0\bar K^0)=(34.3\pm 0.7)\%
\end{array}
\right\}
\Longrightarrow R_{\rm exp}=1.43\pm 0.04\ .
\end{equation}
The current PDG edition \cite{pdg98}, now including the above VEPP-2M data, 
quotes
\begin{equation}
\label{expR}
\left.
\begin{array}{l}
{\rm BR}(\phi\rightarrow K^+K^-)=(49.1\pm 0.8)\%\\[2ex]
{\rm BR}(\phi\rightarrow K^0\bar K^0)=(34.1\pm 0.6)\%
\end{array}
\right\}
\Longrightarrow R_{\rm exp}=1.44\pm 0.04\ ,
\end{equation}
as a result of a global fit, which appears as a very stable result, 
established with a 3\% error.
In the same PDG edition, one can also find $R_{\rm direct} = 1.35\pm 0.06$, 
as the averaged result of the various experiments measuring the ratio 
$\phi\rightarrow K^+K^-/K^0\bar K^0$ directly.  
A reduction of these errors can be expected from DA$\Phi$NE,
where the KLOE experiment has already collected $8\times 10^6\ \phi$-mesons. 
Like in the case of the CMD-2 detector, all the main decay modes of the 
$\phi$ will be measured by the same apparatus and this could bring the 
systematic errors to a minimum, while the statistics will allow to bring 
the statistical error well below the 1\% level. 
Our discussion centers around this ratio $R$ and the possible interest in
studying it with a much reduced experimental error.

%\section{Na\"{\i}ve expectations} 

We shall approach this discussion by starting with the most na\"{\i}ve 
result for the above ratio $R$, {\it i.e.}~$R=1$, which follows from 
assuming that these $\phi \to K \bar{K}$ decay modes proceed exclusively 
via the strong interaction dynamics in the good isospin limit $m_u = m_d$ 
and ignoring phase space differences.
The mass difference between neutral and charged kaons 
---which includes both isospin breaking effects ($m_u \ne m_d$) and 
electromagnetic (photonic) contributions--- considerably increases this 
too-na\"{\i}ve prediction via the (purely kinematical) phase-space factor.
Assuming now perfect isospin symmetry only for the strong interaction dynamics 
(equal couplings for $\phi K^+K^-$ and $\phi K^0 \bar{K}^0$) and knowing 
that $\phi \to K \bar{K}$ are $P$-wave decay modes of a narrow resonance, 
one necessarily has  
\begin{equation}
\label{psR}
R=\frac{\left(1-\frac{4m_{K^+}^2}{M_\phi^2}\right)^{3/2}}
       {\left(1-\frac{4m_{K^0}^2}{M_\phi^2}\right)^{3/2}}=1.528\ ,
\end{equation}
with negligible errors coming from the mass values quoted in the PDG. 
The phase-space correction thus pushes the ratio $R$ two standard 
deviations above its experimental value (\ref{expR}). 
This kinematical correction is exceptionally large because of the vicinity 
of the $\phi$ mass to the $K \bar{K}$ thresholds, which translates into 
considerably large differences between the charged and neutral kaon
momenta 
(or velocities, $v_+ /v_0 = 0.249 / 0.216 = 1.152$), a difference which is 
further increased to its third power in such $P$-wave decay modes. 

This two-$\sigma$ 
discrepancy between experiments and the theoretical tree level predictions 
obviously claims for further corrections.
The most immediate of such corrections is due to electromagnetic radiative 
effects on the ratio $R$, which affect the numerator but not the denominator, 
and which will be discussed in the next section.

\section{Electromagnetic radiative corrections}
\label{emradcor}

Electromagnetic radiative corrections are frequently ignored when dealing 
with strong decays. 
In our case, they could be relevant since, although small, they affect the 
charged decay mode but not the neutral one, and, in order to solve the 
discrepancy in the ratio $R$ under consideration, only a few per cent 
correction is needed. 
Many  years ago they were already considered by Cremmer and Gourdin
\cite{cg} who found a positive correction of the order of 4\% to the 
prediction in Eq.~(\ref{psR}), thus enlarging that discrepancy. 
The dominant contribution was found to arise from the so-called Coulomb
term which is positive for $\phi \to K^+K^-$ and  rather large because of 
the small kaon velocities $v_{\pm} = 0.249$.
A similar increase of the ratio $R$ (some 5\%) by radiative corrections is 
expected by the experimentalists at VEPP-2M \cite{AKH95}, whose quoted
result is inclusive of any vertex correction. 
If we include this correction in the theoretically predicted ratio, the
final result for the radiatively  corrected ratio is then 
$R \simeq 1.59$ \cite{cg}, in agreement with still another independent 
analysis by  Pilkuhn leading to $R$ in the range 
$1.52$--$1.61$ \cite{pilkuhn}. 
To better qualify these statements, we shall now examine in detail
the contribution of such corrections to the ratio $R$.

We have recalculated the electromagnetic radiative corrections to 
$\phi \to K^+K^-$ along the lines of Ref.~\cite{cg}.
For the charged amplitude we start with the usual and simplest tree level 
expression $A_0(\phi \to K^+K^-) = g_0 \epsilon_\mu (p_+ - p_-)^\mu $, 
where $g_0$ is the uncorrected strong coupling constant for 
$\phi K \bar K$, $\epsilon_\mu$ is the $\phi$ polarization and $p_{\pm}$
are the kaon four-momenta.
As is well known, the various contributions to the radiative corrections
can be grouped in two parts. The first part comprises one-loop corrections 
to the uncorrected amplitude $A_0(\phi \to K^+K^-)$. 
This part contains three vertex diagrams with one virtual photon exchanged 
between the two charged-kaons or between the $\phi K^+K^-$ vertex and each
charged-kaon.
In addition, it also contains wave-function renormalization of external 
kaon lines that render the whole amplitude ultraviolet finite\footnote{
Notice that Eq.~(19) in Ref.~\cite{cg} contains a small imaginary part 
while it is supposed to be the real part of the one-loop amplitude.}.
The second part is needed to cancel the infrared divergence. 
It contains three real-photon emission diagrams which are order
$\sqrt\alpha$.
Adding these two parts we find the complete order $\alpha$ corrective 
factor to the $\phi \to K^+K^-$ decay width
\begin{equation}
\label{rc}
\begin{array}{l}
1+C_f+\beta_f \log{{2\Delta E}\over{m_{K^+}}}\equiv
1+\frac{\alpha}{\pi}
\left\{\frac{1+v^2}{2v}\pi^2-2\left(1+\log\frac{2\Delta E}{m_{K^+}}\right)
\left(1+\frac{1+v^2}{2v}\log\frac{1-v}{1+v}\right)\right.\\[2ex]
\ \ \ \ \ \ -\frac{1}{v}\log\frac{1-v}{1+v}
-\frac{1+v^2}{4v}\log\frac{1-v}{1+v}\log\frac{1-v^2}{4}-\frac{1+v^2}{2v}
\left[{\rm Li}_2\left(\frac{2v}{1+v}\right)
     -{\rm Li}_2\left(\frac{-2v}{1-v}\right)\right]\\[2ex]
\ \ \ \ \ \ \left.+\frac{1+v^2}{2v}
\left[{\rm Li}_2\left(\frac{1+v}{2}\right)
     -{\rm Li}_2\left(\frac{1-v}{2}\right)\right]
-\frac{1+v^2}{v}
\left[{\rm Li}_2(v)-{\rm Li}_2(-v)\right]\right\}\ ,
\end{array}
\end{equation}
where $v=\sqrt{1-4 m^2_{K^+}/M^2_\phi}$ is the kaon velocity and 
$\Delta E$ stays for the photon energy resolution. 
For $\Delta E = 1$ MeV the correction (\ref{rc}) amounts to a 4.2\%
increase.
Taking for $\Delta E$ the maximal available photon energy (32.1 MeV, not
far from the energy resolution in the KLOE detector at DA$\Phi$NE, which
is $\approx$ 20 MeV) makes no substantial difference as the main 
contribution comes from the Coulomb term, the first one inside the brackets.

The above discussion ignores the fact that what is actually measured at
VEPP-2M and at DA$\Phi$NE is the ratio 
\begin{equation}
\label{Rsigma}
R_{e^+e^-} \equiv 
{{\sigma(e^+e^-\rightarrow \phi\rightarrow K^+K^-)}\over
 {\sigma(e^+e^-\rightarrow \phi\rightarrow K^0{\bar K^ 0})}}\ ,
\end{equation}
and that radiative corrections to $R$ correspond to consider the ratio of 
the radiatively corrected cross-sections which appear at the numerator and 
denominator of $R_{e^+e^-}$.
In addition to consider both initial and final state corrections, 
a complete treatment also requires to discuss the presence of the $\phi$ 
resonance and the associated distortion of the cross-sections \cite{pancheri}.
At the numerator, radiative corrections include virtual corrections 
as well as emission of soft unobserved photons,
both from the initial and final states, with no interference between
initial and final state radiation for an inclusive measurement 
({\it i.e.}~in a measurement that does not distinguish the charges of the 
kaons) \cite{CE70}.
For the cross-section at the denominator, there are only initial state
radiative corrections since the final kaons are neutral. 
In the absence of final state radiation,
the presence of a narrow resonance like the $\phi$ in the intermediate state
introduces large double logarithms which can be resummed \cite{pancheri,cdg} 
and factorized in an expression like
\begin{equation}
\left({{\Gamma_{\phi}}\over{M_{\phi}}}\right)^{\beta_i} (1+C_i)\ ,
\end{equation}
where $\beta_i={{2\alpha}\over{\pi}}\left(\log{s \over m_e^2}-1\right)$
is the initial state radiation factor and $C_i$ is the finite part of the 
initial virtual and soft photon corrections, which survives after the 
cancellation of the infrared divergence and the exponentiation of the
large resonant dependent factors.
The same factor for initial state radiation appears both at numerator and 
denominator, and since there is no interference between initial and final 
state radiation, the real soft-photon radiative corrections to the initial 
state cancel out in the ratio (\ref{Rsigma}). 
In principle, one should also resum the contributions coming from final
state radiation but the final state radiative factor
$\beta_f={{2\alpha}\over{\pi}}
\left(\frac{1+v^2}{2v}\log\frac{1+v}{1-v}-1\right)\simeq
3.9\times 10^{-4}$ 
is very small and resummation in this case is irrelevant. 
One then obtains the following expression for the ratio $R_{e^+e^-}$ 
as defined in Eq.~(\ref{Rsigma}):
\begin{equation}
\label{rcorrsigma}
R_{e^+e^-} = 
{{\Gamma(\phi\rightarrow K^+K^-)}
\over{\Gamma(\phi\rightarrow K^0\bar K^0)}}\,
{{1+C_i+C_f+\beta_f \log{{2\Delta E}\over{m_{K^+}}}}
\over{1+C_i}}\ .
\end{equation}
Since $C_i\approx {\alpha\over\pi}
\left({3 \over 2}\log{s \over m_e^2}+{\pi^2 \over 3}-2\right)
\simeq 5.6\times 10^{-2}$ \cite{greco&berends}, 
one can expand the denominator in Eq.~(\ref{rcorrsigma}), canceling 
the $C_i$ term and remaining with the final state correction terms $C_f$ 
and $\beta_f$ given explicitly in Eq.~(\ref{rc}). 
We thus conclude that one is justified in using the expressions as above
and that the conventional treatment of radiative corrections increases the 
previous two-$\sigma$ discrepancy between experiment and theory for the 
ratio $R$ to the level of three standard deviations.
 
\section{$SU(2)$-breaking in $\phi K \bar{K}$ vertices}

The $\phi K^+ K^-$ and $\phi K^0 \bar K^0$ vertices are not equal 
(and thus do not cancel in the ratio $R$) 
once $SU(2)$-breaking  effects are taken into account. 
The way $SU(2)$-breaking is usually introduced in the effective lagrangians 
is the same as for $SU(3)$-breaking, namely, via quark mass differences.
In the latter $SU(3)$ case, an improved description of the vector-meson 
couplings to two pseudoscalar-mesons can easily be achieved as shown, 
for example, in Refs.~\cite{bgp,su3}. 
But the situation is by far less convincing when turning to the much smaller 
$SU(2)$-breaking  effects \cite{su2}. 
The essential feature ---common to most models--- is that the dynamics of 
these flavour symmetry breakings suppress the creation of heavier $q \bar q$ 
pairs. 
In the $\phi K^+ K^-$ and $\phi K^0 \bar K^0$ vertices, one needs to
produce a $u \bar u$ and a $d \bar d$ pair, respectively. 
Since the latter is heavier, the $\phi \to K^0 \bar K^0$ decay is further 
suppressed and then the ratio $R$ is further increased. 
To be somewhat more precise, we will consider two recent and independent
models dealing quite explicitly with such kind of effects \cite{bgp,bijnens}. 

In the $SU(3)$-breaking treatment of $VP_1P_2$ vertices by Bijnens 
{\it et al.}~\cite{bijnens}, these decays proceed through two independent 
terms containing the relevant vector and pseudoscalar masses ($M_V$ and 
$m_{1,2}$) and thus incorporating quark-mass breaking effects. 
In the notation of Ref.~\cite{bijnens}, to which we refer for details, 
these $VP_1P_2$ couplings are then proportional to 
\begin{equation}
M^2_V \left( g_V + 2\sqrt{2} f_\chi {m^2_1 + m^2_2 \over M^2_V }\right)\ .
\end{equation}
For the $\phi K^+ K^-$ and $\phi K^0 \bar K^0$ coupling constants, the 
uncorrected strong coupling constant $g_0$ becomes, respectively,  
\begin{equation}
\label{bijnens1}
{M^2_\phi \over 2\sqrt2 g_0 f^2} \left( 1+ 4\sqrt{2} {f_\chi \over g_V} 
{m^2_{K^+,K^0} \over M^2_\phi} \right)\ ,
\end{equation}
with the pion decay constant $f \simeq 92$ MeV. One then obtains the ratio
\begin{equation}
\label{bijnens2}
{g_{\phi K^+K^-} \over g_{\phi K^0 \bar{K}^0}} \simeq 
1 + 4 \sqrt{2} {f_\chi \over g_V} 
{m^2_{K^+} - m^2_{K^0}|_{m_u \neq m_d} \over M^2_\phi } \simeq 1.01\ ,
\end{equation}  
where we have used 
$m^2_{K^+} - m^2_{K^0}|_{m_u \neq m_d} \simeq -6\ 10^{-3}$ GeV$^2$ for
the non-photonic kaon mass difference \cite{bm} and the estimate 
${f_\chi \over g_V} \simeq - {1 \over 3}$
obtained in Ref.~\cite{bijnens} when fitting the $\rho \to \pi \pi$ and 
$K^* \to K \pi$ decay widths. 

Similarly, in the independent treatment of $SU(3)$ symmetry breaking 
\cite{bgp}, some relevant $VP_1P_2$  couplings are given by  
\begin{equation}
\begin{array}{rcl}
g_{\rho \pi \pi} &=& \sqrt{2} g\ ,\\[1ex]
g_{\phi K^+ K^-} &=& g_{\phi K^0 \bar K^0} = -g (1+2c_V)(1-c_A)\ , 
\end{array}
\end{equation}
with $c_V \simeq 0.28$ and $c_A \simeq 0.36$ (see Ref.~\cite{bgp} for
notation and details) mimicking the $SU(3)$ mass difference effects 
discussed in the previous approach \cite{bijnens}. 
The transition from $SU(3)$- to $SU(2)$-breaking offers no difficulties.
One now obtains 
\begin{equation}
\label{bgp1}
{g_{\phi K^+K^-} \over g_{\phi K^0 \bar{K}^0}} \simeq 
1 - {m^2_{K^+} - m^2_{K^0}|_{m_u \neq m_d} \over m^2_K -m^2_\pi } c_A  
\simeq 1.01\ .
\end{equation}
As in the approach of Ref.~\cite{bijnens}, these $SU(2)$-breaking
corrections work in the undesired direction and the discrepancy between 
theory and experiment for the ratio $R$ increases by an additional 2\%.

An independent $SU(2)$-breaking effect can arise from $\rho$-$\phi$
mixing. This is both isospin and Zweig-rule violating, and should therefore 
lead to rather tiny corrections. Indeed, in this context one can 
immediately obtain the following relation among coupling 
constants\footnote{
Notice that this isospin relation not only accounts for 
$\rho(770)$--$\phi$ mixing effects but also for those between $\phi$ and
any other higher mass isovector $\rho$-like resonance.}: 
$g_{\phi K^+K^-} - g_{\phi K^0 \bar{K}^0} = g_{\phi \pi^+ \pi^-}$, 
with a small value for the $g_{\phi \pi^+ \pi^-}$ coupling coming from 
the observed smallness of the $\phi \to \pi^+ \pi^-$ branching ratio 
(${\cal O}(10^{-4})$ \cite{pdg98}) in spite of its much larger phase space. 
A more quantitative estimate is now possible thanks to the recent data on 
$e^+ e^- \to \phi \to \pi^+ \pi^-$ coming from VEPP-2M \cite{VEPP}.
These data describe the pion form factor around the $\phi$ peak, 
$F(s \simeq M^2_\phi)$, in terms of the complex parameter $Z$ by the 
expression 
\begin{equation}
F(s) 
\left(1-{Z M_\phi \Gamma_\phi \over M^2_\phi -s -iM_\phi
\Gamma_\phi}\right)\ .
\end{equation}
This $Z$, in turn, can be easily related to $\epsilon_{\phi \rho}$, 
the complex parameter describing the amount of $\rho$-like 
(or $(u \bar{u}- d \bar{d})/\sqrt{2}$) contamination in the $\phi$ wave 
function. One finds 
\begin{equation}
\epsilon_{\phi \rho} \simeq -{f_\phi \over f_\rho}{\Gamma_\phi \over M_\phi} 
F(s=M^2_\phi ) Z\ ,
\end{equation} 
where the first coefficient 
${f_\phi \over f_\rho} \simeq - {3 \over \sqrt2}$ is the well-known ratio of
$\phi$-$\gamma$ to $\rho$-$\gamma$ couplings.
One finally obtains 
\begin{equation}
{g_{\phi K^+K^-} \over g_{\phi K^0 \bar{K}^0}} \simeq 
1 - \sqrt{2} \Re(\epsilon_{\phi \rho}) \simeq 1.001\ ,
\end{equation}
where an average of the values for $Z$ in Ref.~\cite{VEPP} 
and the parametrization of $F(s=M^2_\phi)$ from Ref.~\cite{gp} have been used
in the  final step.
This time the correction is tiny and the accuracy of our estimate is
rather rough, but again it tends to increase the  discrepancy on the ratio $R$. 

\section{Further attempts}  

Since the discrepancy between the theoretical and experimental value for
$R$ remains (or has even been increased by some additional 2\% due to the 
$SU(2)$-breaking effects just discussed), we have tried to improve our 
analysis in different aspects. 
First, we have taken into account that the couplings of photons to kaons, 
rather than being point-like (as assumed in our previous and conventional
treatment of radiative corrections), are known to be vector-meson dominated
\cite{ABinDaphne}. Accordingly, we have redone the calculation performed 
in Sec.~\ref{emradcor} including the corresponding electromagnetic 
(vector-meson dominated) kaon form-factors. 
Now, not only the decay mode $\phi\rightarrow K^+K^-$ can be affected but
also the $\phi\rightarrow K^0\bar K^0$ one due to the $\rho$, $\omega$ and
$\phi$ mass differences. 
For the $\phi\rightarrow K^+K^-$ case, the contribution of the charged kaon 
form-factor modifies the point-like result for 
$\Gamma(\phi\rightarrow K^+K^-)$ by $\approx 2\ 10^{-3}$. For the case of
$\phi\rightarrow K^0\bar K^0$, a vanishing effect will be obtained in the
limit of exact $SU(3)$ symmetry, and a fraction of the preceding one if
$SU(3)$-broken masses are used. In both cases, the effect of kaon
form-factors on real-photon emission diagrams is null. 
So then, the additional net effect of electromagnetic kaon form-factors on
the ratio $R$ leads to a modification of the point-like radiative corrections
result of Sec.~\ref{emradcor} by some per mille and is thus fully negligible.

A second and independent possibility consists in adopting a different 
framework for $VPP$ decays. This is usually done in terms of more general 
effective lagrangians with $VPP$ vertices containing two derivatives of the 
pseudoscalar fields instead of a single one as in our previous discussion. 
The radiative decay $\rho \to \pi^+ \pi^- \gamma$ ---quite similar to the 
processes we are considering--- has been quite recently analyzed in this 
modern context in Ref.~\cite{huber}.  
The two relevant coupling constants ($F_V$ and $G_V$, in the notation of 
Ref.~\cite{ecker1}) and their relative sign can be fixed to the canonical 
values $F_V = 2 G_V = \sqrt2 f_\pi$ \cite{ecker2} thanks to the experimental 
data for $\rho \to \pi^+ \pi^- \gamma$ and other $\rho$ meson 
processes \cite{pdg98}. 
As discussed in Ref.~\cite{huber}, a good description of these data is then 
achieved in terms of an amplitude that coincides with the one previously 
introduced in Ref.~\cite{bcg}, and which originated from the simple 
one-derivative $VPP$ vertices used by Ref.~\cite{cg} as well as in our
recalculation in Sec.~\ref{emradcor}.
In other words, both types of effective lagrangians lead to exactly the same 
real-photon emission amplitudes once the coupling constants are properly fixed. 
This is also true for the other corrections concerning one-loop effects: 
for the canonical value $F_V = 2 G_V$ one reobtains precisely our previous 
expression in Eq.~(\ref{rc}).  

A third attempt includes the effect of final $K\bar K$ rescattering through 
scalar resonances.
It is well known that the charged kaons emitted in 
$\phi\rightarrow K^+K^-$ are always accompanied by soft photons. 
In the case of single photon emission, the $K^+ K^-$ system is found to be 
in a $J^{PC}=0^{++}$ or $2^{++}$ state with an invariant mass just below the 
$\phi$ mass. The presence of the $J^{PC}=0^{++}$ scalar resonances $f_0(980)$ 
and $a_0(980)$, with masses and decay widths that cover the invariant mass 
range of interest (from the $K\bar K$ threshold to the $\phi$ mass) 
\cite{pdg98}, would suggest that rescattering effects could be 
important\footnote{Rescattering effects from
$2^{++}$ states are suppressed because the nearest tensorial resonances, 
$f_2(1270)$ and $a_2(1310)$, are well above the $\phi$ mass \cite{pdg98}.}.
We have computed these rescattering effects through the exchange of the
$f_0$ and $a_0$ using the charged kaon loop model \cite{nus,lucio,close}.
In this model, the $\phi$ decays into a $K^+ K^-$ system that emits a photon
(from the charged kaon internal lines and from the $\phi K^+ K^-$ vertex)
before rescattering into a final $K^+ K^-$ or $K^0\bar K^0$ state through
the propagation of $f_0$ and $a_0$ resonances.
If the emitted soft photon is unobserved, the process 
$\phi\rightarrow K^+K^-(\gamma)\rightarrow f_0/a_0(\gamma)
\rightarrow K^+ K^-(\gamma)$ or $K^0\bar K^0 (\gamma)$ 
contributes to the ratio $R$, both at the numerator and denominator.
In order to calculate these effects, one needs an estimate of the coupling 
constant $g_{SK{\bar K}}$, where $S$ is either the $f_0$ or the $a_0$. 
Recent measurements of the $\phi\rightarrow f_0\gamma$ and $a_0\gamma$
decay modes at VEPP-2M \cite{novo} are consistent with the predictions 
of the charged kaon loop model for values of the above couplings given by
\begin{equation}
\label{novo}
\frac{g^2_{f_0 K\bar K}}{4\pi}= (1.48\pm 0.32)\ {\rm GeV}^2\ ,\ \ \ \ \ \ 
\frac{g^2_{a_0 K\bar K}}{4\pi}= (1.5\pm 0.5)\ {\rm GeV}^2\ .
\end{equation}
We have then found that the contribution of these kaon loops to the
${\rm BR}(\phi\rightarrow K^+K^- (\gamma))$ is ${\cal O}(10^{-7})$, while 
for ${\rm BR}(\phi\rightarrow K^0\bar K^0 (\gamma))$ is ${\cal O}(10^{-9})$.
For charged kaons in the final state, there is an additional contribution
from the interference between the soft-bremsstrahlung and the scalar
amplitudes. This contribution is given by
\begin{equation}
\label{intcon}
\begin{array}{l}
\Gamma_{\rm int}(\phi\rightarrow K^+K^- (\gamma)) = -\frac{4}{3}\,\alpha\,
\frac{g^2_{\phi K^+K^-}}{4\pi}\,\frac{g^2_{f_0 K^+K^-}}{4\pi}\,
\frac{M_\phi}{2\pi^2 m^2_{K^+}}\,\int_0^{\Delta E} d\omega\,\omega\,
\times\\[2ex]
\ \ \ \ \ \ 
\Re\left(I(a,b)\left[\frac{1}{D_{f_0}(m)}+
                     \frac{g^2_{a_0 K^+K^-}}{g^2_{f_0 K^+K^-}}\,
                     \frac{1}{D_{a_0}(m)}
\right]\right)
\left(w+\frac{1-v^2}{2}\,\log\frac{1-w}{1+w}\right)\ ,
\end{array}
\end{equation}
where $I(a,b)$ is the kaon loop integral defined in
Refs.~\cite{lucio,close}
(with $a\equiv M^2_\phi/m^2_{K^+}$ and $b\equiv m^2/m^2_{K^+}$), 
$D_{f_0/a_0}(m)$ are the scalar propagators and
$w=\sqrt{1-4m^2_{K^+}/m^2}$,
with $m=M_\phi\sqrt{1-2\omega/M_\phi}$ being the invariant mass of the 
$K\bar K$ system and $\omega$ the photon energy.
Using the values in Eq.~(\ref{novo}) for the scalar couplings, 
we find that the interference term, which contributes to $R$ only in the
numerator, is positive and ${\cal O}(10^{-5})$,
{\it i.e.}~completely negligible in spite of being the dominant one.

Admittedly, this estimate of the $K{\bar K}$ rescattering effects is model 
dependent and affected by large uncertainties. Before concluding, we would 
thus like to make a few comments on possible variations on the magnitude of 
the scalar coupling constants and the expressions for the scalar propagators 
$D_{f_0/a_0}(m)$ which enter into our evaluation in the preceding paragraph. 
The values of the couplings $g_{SK{\bar K}}$ depend on the nature of the
scalar mesons, {\it i.e.}~whether they are two- or four-quark states, 
or $K{\bar K}$ molecules.
The results of the $K{\bar K}$ molecule model, in addition to the couplings
$g_{SK{\bar K}}$, depend upon the spatial extension of the scalar $K\bar K$ 
bound state, and the predictions for ${\rm BR}(\phi\rightarrow f_0/a_0\gamma)$ 
(for the same $g_{SK{\bar K}}$) are always smaller than in the purely 
point-like case, {\it i.e.}~the effects on $R$ tend to vanish for more 
extended objects \cite{close}.
The two-quark model, irrespectively of the $s \bar{s}$ 
{\it vs.}~$(u \bar{u} + d \bar{d})/\sqrt{2}$ quark content of the $f_0$, 
predicts too small values (see, for example, Refs.~\cite{close,achasov})
for the branching ratios ${\rm BR}(\phi\rightarrow f_0/a_0\gamma)$ \cite{novo}, 
and is unable anyway to account for the near mass degeneracy of the isoscalar 
$f_0$ and isovector $a_0$. On the other hand, such mass degeneracy is well 
understood in the four-quark model, critically reexamined very recently in
Refs.~\cite{jaffe,schechter}. The four-quark model also predicts values for 
$g_{SK{\bar K}}$ that seem to be in agreement with the available measurements 
of ${\rm BR}(\phi\rightarrow f_0/a_0\gamma)$ \cite{novo,achasov}.
In all cases, the different possibilities are found to modify the previously 
quoted sizes of the $K{\bar K}$ rescattering effects by at most one order of
magnitude. Something similar happens with the lack of consensus on the 
specific form for the scalar propagators to be used in these estimates. 
Here the uncertainties arise because of the opening of the $K{\bar K}$ 
channels quite close to the nominal scalar masses. This translates into sharp 
modifications of the conventional Breit-Wigner curves and changes the size of 
the $K{\bar K}$ rescattering effects again by one order of magnitude. 
Although affected by large uncertainties, the contributions coming from 
final-state $K \bar {K}$ rescattering are thus found to be negligible and 
their effects on the ratio $R$ irrelevant.

\section{Conclusions}

In this letter, we have performed a discussion of the ratio 
$R \equiv \phi \to K^+ K^- /$ $\!\!K^0 \bar{K^0}$. 
From the experimental point of view, the value $R_{\rm exp} = 1.44 \pm 0.04$ 
seems to be firmly established \cite{pdg98}. However, in our present 
theoretical analysis of this ratio $R$ we have failed to reproduce the value 
$R_{\rm exp}$ quoted above. 
In a first and conservative attempt, 
including isospin symmetry for the strong vertices and the appropriate 
phase-space factor, one obtains $R = 1.53$ which is two $\sigma$'s above 
$R_{\rm exp}$.  
In a second step, we have also included conventional electromagnetic
radiative corrections to order $\alpha$, thus obtaining $R = 1.59$ and 
increasing the previous discrepancy up to three $\sigma$'s. 
This value confirms some existing results and has been checked to be quite 
independent from the details of the relevant vertices. 
In a third step, we have tried to correct our predictions for $R$ introducing 
various isospin breaking corrections to the $\phi K\bar K$ coupling constants.
As a result, the ratio $R$ is found to be further increased by some 2\%, 
an estimate affected by rather large errors reflecting our poor knowledge on 
the $SU(2)$-breaking details. 
In view of all this, we have introduced final-state rescattering effects which 
should be dominated by the almost on-shell formation of the $f_0(980)$ and 
$a_0(980)$ resonances in the $S$-wave $K\bar{K}$ channel. 
The controversial nature of these scalar resonances allows for quite disparate 
estimates of their effects, but one can safely conclude that they are well 
below those previously mentioned. 
The disagreement on the ratio $R$ persists well above two (experimental) 
standard deviations.
Higher statistics from DA$\Phi$NE are expected in order to settle 
definitively whether the discrepancy on $R$ is a real problem, or final 
agreement between theory and experimental data can be achieved.

\section*{Acknowledgements}
We would like to thank J.~Bijnens, M.~Block, J.~Gasser, E.~Oset, M.R.~Pennington 
and E.P.~Solodov for helpful comments and clarifying discussions.
Work partly supported by the EEC, TMR-CT98-0169, EURODAPHNE network.
J.L.~Lucio M.~acknowledges partial financial support from CONACyT and
CONCyTEG.

\end{document}